\documentclass{article}
\usepackage{spconf}
\usepackage{epstopdf}
\usepackage[draft]{hyperref} 
\usepackage{setspace}
\usepackage{amsmath}
\usepackage{amssymb}
\usepackage{amsthm}
\usepackage{amsfonts}
\usepackage{color}
\usepackage{graphicx}
\usepackage{subfigure}  
\usepackage{cite}
\usepackage{mathtools}
\usepackage{stmaryrd}
\usepackage[]{algorithmicx}
\usepackage{algorithm}
\usepackage{algpseudocode}
\usepackage{bbm}
\usepackage[mathscr]{euscript}
\usepackage{mathrsfs}

\usepackage{soul}

\title{Residual Recovery Algorithm For Modulo Sampling}
\name{Eyar Azar, Satish Mulleti, and Yonina C. Eldar\thanks{ This work has received funding from the Benoziyo Endowment Fund for the Advancement of Science, Estate of Olga Klein -– Astrachanthe; European Union’s Horizon 2020 research and innovation program under grant No. 646804-ERC-COG-BNYQ; and the Israel Science Foundation under grant no. 0100101.}}
\address{Faculty of Math and Computer Science, Weizmann Institute of Science, Israel \\
Emails: eyarazar@gmail.com, mulleti.satish@gmail.com, yonina.eldar@weizmann.ac.il}
\begin{document}
\ninept
\maketitle
\begin{abstract}
Two important attributes of analog to digital converters (ADCs) are its sampling rate and dynamic range. The sampling rate should be greater than or equal to the Nyquist rate for bandlimited signals with bounded energy. It is also desired that the signals' dynamic range should be within that of the ADC's; otherwise, the signal will be clipped. A modulo operator has been recently suggested prior to sampling to restrict the dynamic range. Due to the nonlinearity of the modulo operation, the samples are distorted. Existing recovery algorithms to recover the signal from its modulo samples operate at a high sampling rate and are not robust in the presence of noise.
In this paper, we propose a robust algorithm to recover the signal from the modulo samples which operates at lower sampling rate compared to existing techniques. We also show that our method has lower error compared to existing approaches for a given sampling rate, noise level, and dynamic range of the ADC. Our results lead to less constrained hardware design to address dynamic range issues while operating at the lowest rate possible. 
\end{abstract}
\
\section{Introduction}
Analog to digital converters (ADCs) play a crucial role in digital signal processing. The cost and power consumption of ADCs increase with higher sampling rate. Hence, it is desirable to sample at the lowest possible rates \cite{eldar_2015sampling, subnyquist_review, eldar_SI}. Apart from sampling rate, another key attribute of an ADC is its dynamic range. Ideally, the dynamic range of an ADC should be larger than that of the input analog signal; otherwise, the signal is clipped which leads to loss of information.

One approach to address dynamic range  issues is to use a modulo operation prior to sampling where the input signal is folded back when it crosses the dynamic range of the ADC. Hardware realizations of such high-dynamic range ADCs, also known as \emph{self-reset} ADCs, are discussed in the context of imaging in \cite{sradc_park, sradc_sasagwa, sradc_yuan, krishna2019unlimited}. Along with the samples of the modulo signal, these architectures also store side information such as the amount of folding for each sample, or, the sign of the folding. Measuring the side information leads to complex circuitry at the sampler but enables computationally simple recovery of the signal from the folded samples.

Bhandari et al. considered an alternative approach, called \emph{unlimited sampling}, where no side information is  measured and only folded or modulo samples are used for recovery \cite{unlimited_sampling17, uls_tsp}. 
The authors proposed an algorithm to determine the true or unfolded samples from the modulo ones by applying an extension of Itoh's unwrapping algorithm \cite{itoh}. Perfect recovery of bandlimited inputs is guaranteed provided that the sampling rate is greater than or equal to $(2\pi e)$-times the Nyquist rate where $e$ is the Euler's constant \cite{unlimited_sampling17, uls_tsp}. The recovery algorithm requires very high sampling rate and is sensitive to noise due to the use of higher-order differences.


Romanov and Ordentlich \cite{uls_romonov} improved upon these  results and proposed an algorithm that requires a sampling rate slightly above the Nyquist rate. The authors leverage the fact that there exists a time instant beyond which the signal lies within the dynamic range of the ADC. From these unfolded samples, the folded samples are predicted by using the correlation among the samples of bandlimited inputs. However, simulation results of the algorithm are not presented, especially, in the presence of noise. GAN and Liu \cite{uls_multichannel} considered a multichannel extension of modulo sampling. In the absence of noise, the authors showed that two channels, each of them operating at the Nyquist rate, are sufficient to undo the modulo operation provided that the dynamic ranges of the two ADCs are coprime. Reconstruction is based on the application of the Chinese remainder theorem. In the presence of noise, more than two channels are considered and a lattice-based optimization method is used for recovery. 
Modulo sampling has been extended to different problems and signal models such as periodic bandlimited signals \cite{bhandari2021unlimited}, wavelets \cite{uls_wavelet}, mixture of sinusoids \cite{uls_sinmix}, finite-rate-of-innovation signals \cite{uls_fri}, multi-dimensional signals \cite{uls_md}, sparse vector recovery \cite{uls_gamp, uls_sparsevec}, direction of arrival estimation \cite{uls_doa}, computed tomography \cite{uls_radon}, and graph signals \cite{uls_graph}.

 
 In this paper, we consider recovering unfolded samples of a bandlimited input from modulo samples and propose a sampling efficient and robust algorithm. Our method relies on the fact that the residual signal, that is, the difference between the true signal and the output of the nonlinear operator, is time-limited for bandlimited signals. Hence, the discrete-time Fourier transform (DTFT) of the residual signal is a trigonometric polynomial. Furthermore, while the true signal is bandlimited the folded signal has spectral content beyond the bandwidth. Therefore by slightly oversampling, that is, by sampling slightly above the Nyquist rate at the output of the nonlinear operator, we can recover the partial DTFT of the residual signal over a frequency interval from the modulo samples. Using this partial DTFT, we determine the residual sequence by fitting the trigonometric polynomial. We pose the estimation of the residual signal from the partial DTFT as a constrained optimization problem and use a projected gradient decent (PGD) method to solve it. 
 We compare our algorithm with those in \cite{uls_tsp} and \cite{uls_romonov} and show that the proposed method can reconstruct the signal at a lower sampling rate for a given noise level and dynamic range of the ADC.

The rest of the paper is organized as follows. In Section~\ref{sec:problem_formulation}, we formulate the problem.  
In Section \ref{sec:algo} we present the proposed algorithm. 
Simulation results are provided in Section~\ref{sec:simulation} followed by conclusions. 

We use the following notations throughout the paper. 
Uniform samples of $f(t)$ are denoted by $f(nT_s)$ where $T_s >0$ is the sampling interval and $n \in \mathbb{Z}$. The corresponding sampling rate is the $\omega_s = \frac{2\pi}{T_s}$ rads/sec. For a sequence $f(nT_s)$ its corresponding boldfaced symbol $\mathbf{f}$ denotes its vector form where the $n$-th entry is given by $\mathbf{f}[n] = f(nT_s)$. Operator $\mathcal{F}$ denotes the DTFT: 
\begin{align}
    \label{eq:DTFT}
    \mathcal{F}\mathbf{f}= F(e^{\mathrm{j}\omega T_s}) = \sum_{n \in \mathbb{Z}}f(nT_s)\, e^{-\mathrm{j}\omega nT_s}
\end{align}
For any interval $\rho \subset (-\omega_s/2,\,\omega_s/2)$, $\mathcal{F}_{\rho}\mathbf{f}$ denotes a partial DTFT $F(e^{\mathrm{j}\omega T_s})$ evaluated over $\omega \in \rho$, and  $\mathcal{F}^*_{\rho}$ denotes the adjoint operator of $\mathcal{F}_{\rho}$. Specifically, we have 
\begin{align}
    \label{eq: adjoint_partiel_DTFT}
    \mathcal{F}^*_{\rho}\mathcal{F}\mathbf{f}[n] = \frac{T_s}{2\pi}\int_\rho {F(e^{\mathrm{j}\omega T_s})\, e^{\mathrm{j}\omega n T_s} \mathrm{d}\omega}, \quad n\in \mathbb{Z}.
\end{align}
For any $N \in \mathbb{N}$, the set $\mathcal{S}_{N}$ denotes sequences that have support over $\{-N, \cdots, N\}$, and $P_{\mathcal{S}_N}$ denotes the orthogonal projection onto the set ${\mathcal{S}_N}$. Specifically, when $P_{\mathcal{S}_N}$ operates on a sequence, it sets all the samples beyond $\{-N, \cdots, N\}$ to zero.
The symbol $\mathrm{B}_{\omega_m}$ denotes the space of analog signals that are bandlimited to the frequency interval $(-\omega_m, \omega_m)$. The Nyquist sampling rate for signals in $\mathrm{B}_{\omega_m}$ is given by $2\omega_m$. For any sampling interval $T_s$, we define the over-sampling factor (OF) as $\frac{\omega_s}{2\omega_m}$.
For any $a \in \mathbb{R}$ and $\lambda \in \mathbb{R}^+$, the modulo operation $\mathcal{M}_{\lambda}(\cdot)$ is given as
	\begin{equation}
	\mathcal{M}_{\lambda}(a) = (a+\lambda)\,\, \text{mod}\,\, 2\lambda -\lambda.	
	\label{eq:mod_def}
	\end{equation}

\begin{figure}
    \centering
    \includegraphics[width = 3 in]{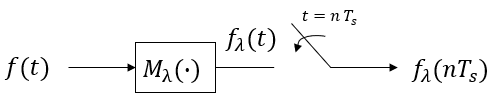}
    \caption{Modulo sampling: The bandlimited signal $f(t)$ is processed through a modulo operator $\mathcal{M}_{\lambda}$ and then sampled by an ADC with sampling interval $T_s$.}
    \label{fig:sampling_block}
\end{figure}


\section{Problem Statement}
\label{sec:problem_formulation}
Consider a BL signal $f(t) \in L^2(\mathbb{R}) \cap  B_{\omega_{m}}$ which is sampled using an ADC with a dynamic range $[-\lambda, \lambda]$. We assume that $||f(t)||_{\infty}> \lambda$ and hence direct sampling results in clipping. To avoid clipping, we first process $f(t)$ through a modulo operator $\mathcal{M}_{\lambda}(\cdot)$ that limits the dynamic range. The output $f_\lambda(t) = \mathcal{M}_{\lambda}(f(t))$ is then sampled as shown in Fig.~\ref{fig:sampling_block}. The sampling rate has to be greater than the Nyquist rate for uniquely identifying $f(t)$ from $f_{\lambda}(nT_s)$ \cite{uls_tsp}.   

Existing algorithms to reconstruct $f(t)$ from modulo samples $f_{\lambda}(nT_s)$ either operate at very high sampling rates or result in large error in the presence of noisy samples. Our objective is to devise a robust algorithm that reconstructs $f(t)$ from its modulo samples $f_{\lambda}(nT_s)$ with lower reconstruction error in the presence of noise. In addition, the algorithm should operate close to the Nyquist rate of $f(t)$. Note that if one recovers the \emph{unfolded} or \emph{true} samples $f(nT_s)$ from $f_{\lambda}(nT_s)$ then $f(t)$ can be uniquely reconstructed from $f(nT_s)$ provided the sampling rate is greater than or equal to the Nyquist rate.

The modulo samples can be written as a linear combination of the true samples and a residual signal:
\begin{align}
    f_\lambda(nT_s) = f(nT_s) + z(nT_s),
    \label{eq:resiudual}
\end{align}
 where the residual sequence has values that are integer multiples of $2\lambda$. Given $f_\lambda(nT_s)$, if one can first estimate $z(nT_s)$, then $f(nT_s)$ can be determined. To this end we use the following two properties of finite energy BL signals.
\begin{itemize}
        \item Time-domain separation \cite{uls_romonov}: From the Riemann-Lebesgue Lemma it can be shown that $\lim_{|t| \to \infty}f(t) = 0$. This implies that for any $\lambda >0$ there exists an integer $N_{\lambda}$ such that $|f(nT_s)| < \lambda, \forall |n|> N_{\lambda}$. Hence, for $|n|> N_{\lambda}$, we have $f_\lambda(nT_s) = f(nT_s)$ and $z(nT_s) = 0$. Thus the modulo samples are equal to the true samples over a set of indices and it can be used to distinguish the residual from the modulo samples in time. 
        \item Fourier-domain separation: Since the signal is sampled above the Nyquist rate to allow perfect recovery, we have that 
        \begin{equation}
       F(e^{\mathrm{j}\omega T_s}) = 0, \quad \text{for} \quad  \omega_m < |\omega| < \omega_s / 2.
    \label{eq: zero_dtft0}
    \end{equation}
    By using the linearity of DTFT, from \eqref{eq:resiudual} we have that
    \begin{equation}
       F_\lambda(e^{\mathrm{j}\omega T_s}) = Z(e^{\mathrm{j}\omega T_s}), \quad \text{for} \quad  \omega_m < |\omega| < \omega_s / 2.
    \label{eq: zero_dtft1}
    \end{equation}
        This implies that one can differentiate the DTFT of the true samples and that of the residual by sampling above the Nyquist rate and looking beyond the bandwidth. 
\end{itemize}
 In rest of the discussion, we assume that $N_{\lambda}$ is known. From the time-domain separation, we infer that the residual signal has finite support on the integer set $\mathcal{N}_{\lambda} = \{-N_\lambda, \cdots, N_{\lambda}\}$. Combining the time-domain and the frequency-domain separations we arrive at the following relation:
 \begin{equation}
            F_{\lambda}(e^{\mathrm{j}\omega T_s}) = \sum_{n={-N_\lambda}}^{N_\lambda}z(n T_s)e^{-\mathrm{j} n T_s\omega},
            \label{eq:combine}
\end{equation}
    for $\omega_m < |\omega| < \omega_s / 2$. Due to its finite support, the DTFT of $z(nT_s)$ is a trigonometric polynomial and it is given over an interval. From \eqref{eq:combine} one can determine $z(nT_s)$ by sampling $F_{\lambda}(e^{\mathrm{j}\omega T_s})$ at $2N_\lambda +1$ points over the interval $\rho = (-\omega_s/2, -\omega_m) \cup (\omega_m, \omega_s/2)$ and inverting the resulting set of linear equations. Using \eqref{eq:resiudual}, the true samples may then be recovery. However, in practice this approach is computationally infeasible for large $N_\lambda$. Even for small $N_\lambda$ this inversion may not be stable in the presence of noise.
    
    In \cite{uls_romonov}, the authors exploit the time-domain separation property together with the fact that the samples $f(nT_s)$ are correlated when sampled above the Nyquist rate. Then the samples beyond the dynamic range of the ADC, $\{f(nT_s)\}_{n = -N_{\lambda}}^{N_{\lambda}}$, are estimated from the remaining samples. In the presence of noise, we show that this approach results in a large error for small $\lambda$. In \cite{uls_tsp}, the authors recover $f(nT_s)$ by using the fact that for very high oversampling factors, higher-order differences of $f_\lambda(nT_s)$ and $f(nT_s)$ are equal in the absence of noise. This approach requires a large amount of over-sampling and is not stable in the presence of noise due to the use of high-order differences. 
    
    In our approach, we combine both properties  of separation in time and frequency to determine the residual sequence. Specifically, we propose a stable, lowrate, and computationally feasible algorithm to estimate $z(nT_s)$ from $F_{\lambda}(e^{\mathrm{j}\omega T_s})$, relying on  \eqref{eq:combine}.

    \section{A Robust and Lowrate Recovery Algorithm}
\label{sec:algo}
To estimate $z(nT_s)$ from $F_{\lambda}(e^{\mathrm{j}\omega T_s})$ we consider an optimization framework . By using the operator and vector notations we rewrite \eqref{eq: zero_dtft1} as 
\begin{align}
\mathcal{F}_{\rho}\mathbf{f}_{\lambda} =  \mathcal{F}_{\rho}\mathbf{z},
\label{eq:equal_vector}
\end{align}
where $\rho = (-\omega_s/2, -\omega_m) \cup (\omega_m, \omega_s/2)$. Since $z(nT_s)$ is time-limited to $\mathcal{N}_{\lambda}$, we have that 
\begin{align}
   \mathbf{z} \in \mathcal{S}_{N_\lambda}.
   \label{eq:constarint}
\end{align}
Given the data-fitting term in \eqref{eq:equal_vector} and support constraint, recovery of $\mathbf{z}$ can be written as the following optimization problem:
\begin{align}
    \underset{\mathbf{z}}{\min} \quad \mathrm{C}(\mathbf{z}) = \dfrac{1}{2}\|\mathcal{F}_{\rho}\mathbf{f}_{\lambda} - \mathcal{F}_{\rho}\mathbf{z}\|^2 \quad \text{s.t.} \quad \mathbf{z} \in \mathcal{S}_{N_\lambda}.
    \label{eq:opt2}
\end{align}

Problem \eqref{eq:opt2} can be solved using a PGD method where at each iteration the solution iterates towards the negative gradient of the cost $\mathrm{C}(\mathbf{z})$ and is then projected onto the set $\mathcal{S}_{N_\lambda}$. In summary, starting from an initial point $\mathbf{z}^0 \in \mathcal{S}_{N_\lambda}$, the steps at the $k$-th iteration are
\begin{equation}
 \begin{split}
        \mathbf{y}^{k} &= \mathbf{z}^{k-1} - \gamma_k \nabla \mathrm{C}(\mathbf{z}^{k-1}) \\
    \mathbf{z}^k &= P_{\mathcal{S}_{N_{\lambda}}}(\mathbf{y}^k),
     \end{split}
     \label{eq:pgd}
\end{equation}
  where $\gamma_k >0$ is a suitable step-size, $\nabla \mathrm{C}(\mathbf{z}) = \mathcal{F}^*_{\rho}\mathcal{F}_{\rho}{(\mathbf{z} - \mathbf{f}_{\lambda})}$ is the gradient of $\mathrm{C(\mathbf{z})}$ and $P_{S_{N_\lambda}}(\mathbf{y})$ is the orthogonal projection onto $\mathcal{S}_{N_{\lambda}}$. The operator $\mathcal{F}^*_{\rho}\mathcal{F}_{\rho}$ is a highpass operation. The sequence $\mathcal{F}^*_{\rho}\mathcal{F}_{\rho}(\mathbf{z} - \mathbf{f}_{\lambda})$ can be computed by convolving the sequence $(\mathbf{z} - \mathbf{f}_{\lambda})$ with an ideal highpass filter with spectral support over $\rho$.
    In addition to the support constraint, we know that every element of $\mathbf{z}$ is in $\in 2\lambda \mathbb{Z}$ \cite{unlimited_sampling17}. To improve the estimation accuracy, we use this property to round the solution of the optimization problem to a nearest integer multiple of $2\lambda$.  
    
    While the rounding operation followed by PGD gives a good recovery of $\mathbf{z}$ from the modulo samples for small $N_{\lambda}$, for large $N_{\lambda}$ the estimation 
    tends to be more accurate at the edges of the support. 
    Using this observation, we propose a sequential approach to improve the accuracy of estimation of the remaining samples, as detailed next.
    Starting from a given $N_\lambda$, let the projection gradient descent algorithm estimate of $\mathbf{z}$ be $\mathbf{\hat{z}}$. The estimate has support over $\mathcal{N}_\lambda$ and its values are integer multiples of $2\lambda$. In the absence of noise, 
    \begin{align}
        \mathbf{z}[n] = \mathbf{\hat{z}}[n], \quad n = \pm N_\lambda.
        \label{eq:edge_samples}
    \end{align}
    To estimate the remaining samples of $\mathbf{z}$ accurately, we define another sequence as 
    \begin{align}
      \label{eq:rec_signal}
      \mathbf{\hat{f}}_\lambda = \mathbf{f_\lambda} - \mathbf{\hat{z}}.
    \end{align}
    Combining \eqref{eq:resiudual} and \eqref{eq:rec_signal} we have
    \begin{align}
    \label{eq:new_prob}
       \mathbf{\hat{f}}_\lambda  = \mathbf{f} +(\mathbf{z} - \mathbf{\hat{z}}).
    \end{align}
    From \eqref{eq:edge_samples} and \eqref{eq:new_prob} we have that $\mathbf{\hat{f}}[n] = \mathbf{f}[n],\, |n|>N_\lambda-1$. As a result, the new residual sequence $(\mathbf{z} - \mathbf{\hat{z}})$ has support over $\{-(N_\lambda-1), \cdots, (N_\lambda-1)\}$, that is, $(\mathbf{z} - \mathbf{\hat{z}}) \in \mathcal{S}_{N_\lambda -1}$ and $(\mathbf{z} - \mathbf{\hat{z}}) \in 2\lambda \mathbb{Z}$. Hence, $\mathbf{\hat{f}}_{\lambda}$ has a similar decomposition as in \eqref{eq:resiudual} except for the fact that its values need not be in the range $[-\lambda, \lambda]$. Despite that, we can redefine the optimization problem as in \eqref{eq:opt2} to estimate $(\mathbf{z} - \mathbf{\hat{z}})$ from $\mathbf{\hat{f}}_\lambda$ and use the PGD iterations as in \eqref{eq:pgd} to solve it. The residue $(\mathbf{z} - \mathbf{\hat{z}})$ is correctly estimated for $n = \pm (N_\lambda-1)$, from which $\mathbf{f}$ can be determined at those locations. The process is repeated until all the samples are estimated. Since our algorithm uses the information of the residual beyond the bandwidth of the signal we denote it as \emph{beyond bandwidth residual reconstruction ($B^2R^2$)} algorithm and summarize it in Algorithm~\ref{alg:algorithm_new}. 
    
    For initialization, one can set $\mathbf{z}^0$ as $P_{\mathcal{S}_{N_{\lambda}}}\{\mathcal{F}^*_{\rho}\mathcal{F}_{\rho} \mathbf{f}_{\lambda}\}$. This is inverse-partial DTFT of $\mathcal{F}_\rho\mathbf{z}$ and we found that it serves as a good initial point.

\begin{algorithm}[!t]
  \caption{$B^2R^2$ for recovery of BL signals from modulo samples}\label{euclid}\label{alg:algorithm_new}
  \begin{algorithmic}[1]
     \State {\bf Input}$f_{\lambda}(n T_s)$ or $\mathbf{f}_{\lambda}$, $\lambda$, $\rho$ and $N_{\lambda}$
    \State \textbf{Intialize:} $\mathbf{\hat{f}}_{\lambda} = \mathbf{f_\lambda}$, $\mathbf{z}^0 \in S_{N_\lambda}$ 
      \While{$N_\lambda >0$} 
         \For{$k = 1$, $k{+}{+}$, Until stopping criteria}
       \State Choose step size $\gamma_k$ by backtracking line search
            \State $\mathbf{y}^k = \mathbf{z}^{k-1} - \gamma_k \mathcal{F^*}_{\rho}\mathcal{F}_{\rho}{(\mathbf{z}^{k-1} -\mathbf{\hat{f}}_{\lambda})}$
            \State $\mathbf{z}^k = P_{S_{N_\lambda}}(\mathbf{y^k})$
         \EndFor
         \State $\mathbf{\hat{z}} = \mathbf{z}^k$, \Comment{Estimation after applying PGD algorithm}
        \State $\mathbf{\hat{z}} \leftarrow \left \lceil \frac{\lfloor \mathbf{\hat{z}} / \lambda \rfloor}{2} \right \rceil $  \Comment{rounding to $2\lambda \mathbb{Z}$},
        \State $\mathbf{\hat{f}}_{\lambda} \leftarrow \mathbf{\hat{f}}_{\lambda} - \mathbf{\hat{z}}$
        \State $N_{\lambda} \leftarrow N_{\lambda}-1$
        \State $\mathbf{z}^0 =  P_{S_{N_\lambda}}(\mathbf{\hat{z}})$ 
      \EndWhile\label{euclidendwhile}
      \Statex \textbf{Output:}
        $\mathbf{f} = \mathbf{\hat{f}}_{\lambda}$
  \end{algorithmic}
\end{algorithm}

\section{Simulations Results}
\label{sec:simulation}
    In this section we present numerical results of different methods for recovery of a real bandlimited signal from its modulo samples. We consider noisy samples and assess the performance as a function of OF and $\lambda$. Specifically, we compare the $B^2 R^2$ algorithm with the \emph{higher-order differences} approach \cite{unlimited_sampling17, uls_tsp} and \emph{Chebyshev polynomial filter} method \cite{uls_romonov}. 
    We consider an additive noise model  
        \begin{align}
            \tilde{f}_{\lambda}(n T_s) = {f}_{\lambda}(n T_s) + v(n T_s) = \mathcal{M}_{\lambda}f(n T_s) + v(n T_s), \nonumber
        \label{eq: noisy_samples}
        \end{align}
        where $v(n T_s)$ denotes noise. Throughout the experiments we normalize the bandlimited signals to have maximum amplitude of one. In the simulations SNR is computed as 
        \begin{align}
            \text{SNR} = 10 \log\left(\frac{\frac{1}{N}||f_{\lambda}(n T_s)||^2}{\sigma_v ^2}\right), \nonumber
        \end{align}
        where $N$ is the length of $\tilde{f}_{\lambda}(n T_s)$ and $\sigma_v ^2$ is the variance of $v(nT_s)$. To compare the reconstruction accuracy of different algorithms we consider the normalized mean-squared error (MSE) as 
        \begin{align}
             \text{MSE} = \frac{||f(nT_s) - \hat{f}(n Ts)||^2}{ ||f(nT_s)||^2},
        \end{align}
        where $\hat{f}(n T_s)$ denotes the estimate of $f(n T_s)$. For each noise level, we consider 250 independent noise realizations and average the MSE over them. In these experiments, we assume that the noise samples $v(nT_s)$ are independent and identically distributed Gaussian random variables with zero mean. The variance is set to achieve a desired SNR. 
      
            \begin{figure}[!h]
                \centering
                \includegraphics[width=3.3in]{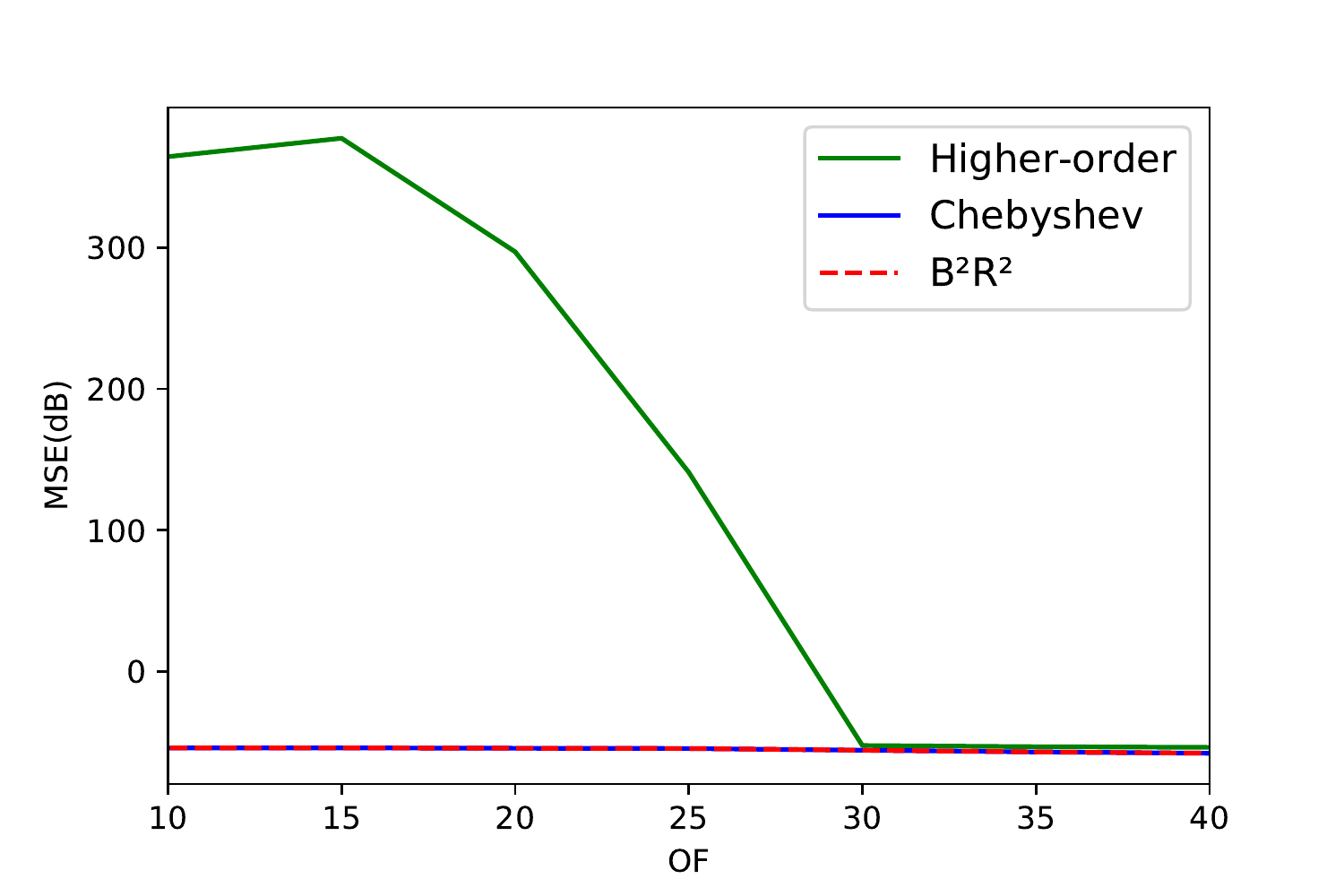}
                \caption{ Comparison of algorithms in terms of MSE in recovering a bandlimited signal from modulo samples with  $\lambda$ = 0.025, and SNR = 25 dB; The higher-order difference approach has error of $-60$ dB for OF $\geq 30$ whereas the remaining methods are able to achieve $-60$ dB error for OF  $= 10$.}
            \label{fig:high_OF}
            \end{figure}
            \begin{figure}
                \centering
               \includegraphics[width=3.4in]{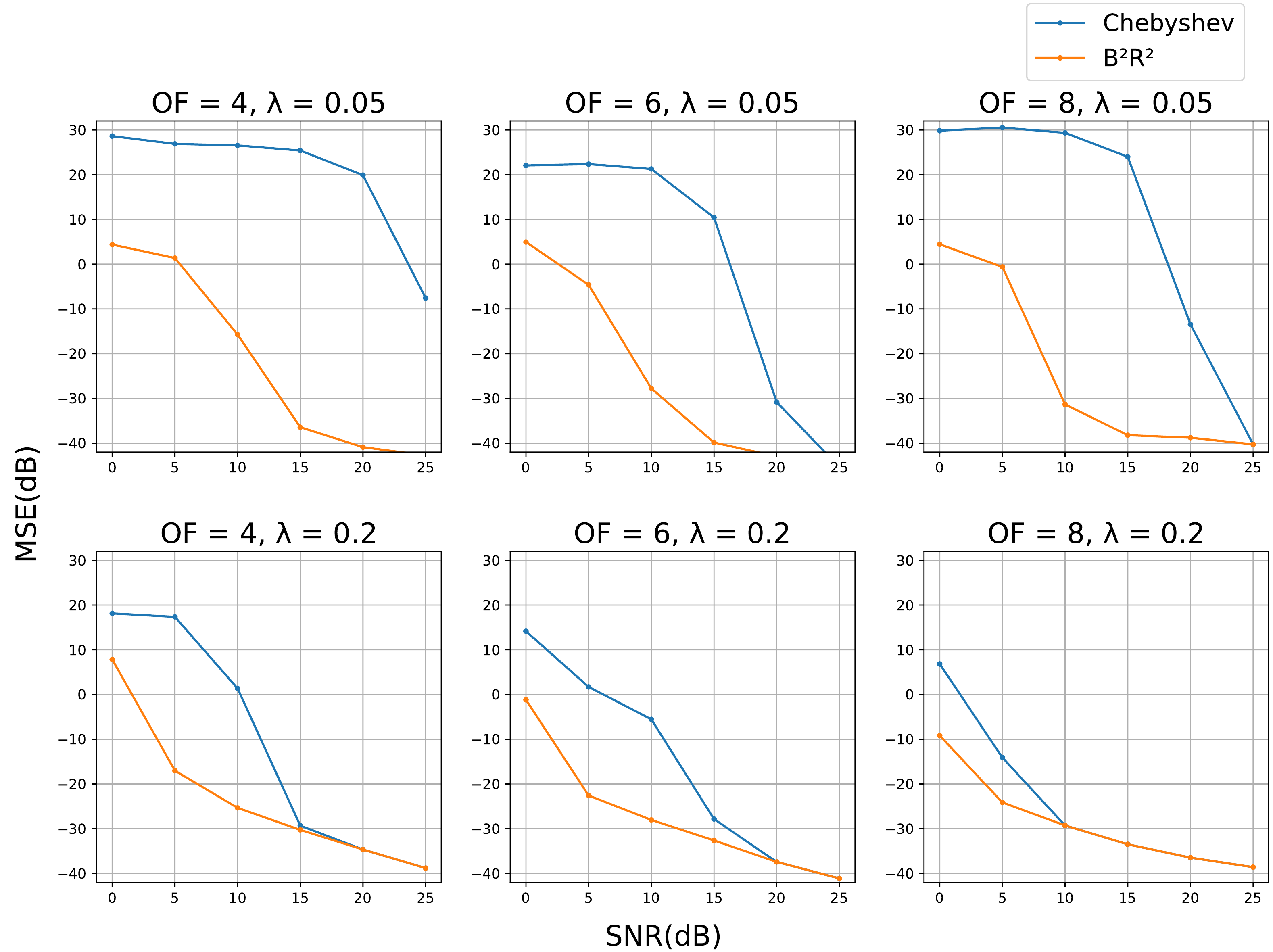}
                \caption{ Comparison of $B^2R^2$ and Chebyshev algorithms in terms of MSE in recovering a bandlimited signal from modulo samples with OF = 4, 6, 8, and  $\lambda$ = 0.05, 0.2.}
            \label{fig:comparesion}
            \end{figure}
            
            \begin{figure}[!h]
                \centering
               \includegraphics[width=3.4in]{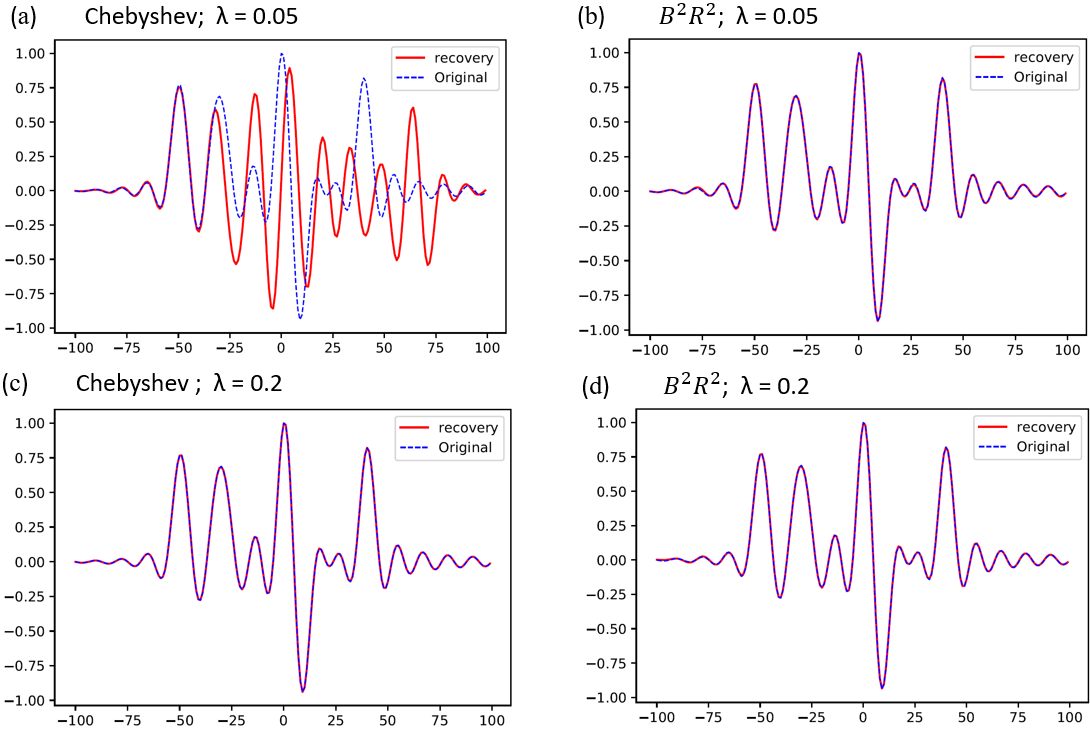}
                \caption{ Comparison of $B^2R^2$ and Chebyshev algorithms in terms of MSE in recovering a bandlimited signal from modulo samples with OF = 6,  $\lambda$ = 0.05, 0.2, and SNR = 15 dB. }
            \label{fig:example}
            \end{figure}
            
         In the first experiment, we compare the methods for different OFs with $\lambda = 0.025$ and SNR of 25 dB. The MSEs for the three methods are shown in Fig.~\ref{fig:high_OF}. We observe that the \emph{higher-order difference} approach is able to reconstruct signals up to an error of $-60$ dB for OF$\geq 30$ whereas the rest of the two approaches achieve $-60$ dB error for OF  = 10. In the presence of noise, in order to ensure that the higher-order difference of the samples of the BL signal are bounded by $2\lambda$, a large amount of OF is required for small values $\lambda$, whereas both $B^2R^2$ and Chebyshev approaches are independent of higher-order differences and require lower oversampling.
         
         Given that the \emph{higher-order differences} method requires large OF for small $\lambda$ or low SNR, next, we compare the proposed $B^2R^2$ with the Chebyshev method. Fig.~\ref{fig:comparesion} shows the MSE as function of SNR of the $B^2 R^2$ and Chebyshev algorithms for $\lambda = 0.05,0.2$ and $\text{OF} = 4,6,8$.
         We note that as the SNR and OF increase, the MSEs of the algorithms decrease. 
         In addition, we notice that when $\lambda$ decreases from $0.2$ to $0.05$, there are more foldings of the samples and it is not easy to reconstruct the signals. 
         The proposed algorithm incurs lower error for a given OF and SNR, especially for a small value of $\lambda$. To illustrate this further, in Fig. 4 we show an examples of recovery of both methods for $\lambda = 0.05, 0.2$ for SNR = 15 dB and OF = 6. We note that while both methods are able to recover the signal accurately for $\lambda = 0.2$, the Chebyshev method fails to converge for $\lambda = 0.05$.


\section{conclusion}
A low dynamic range of an ADC limits its application to bounded signals. A modulo operation is used to address the dynamic range issue, but induces nonlinearity in the samples. Existing algorithms to recover the signal from modulo samples either operate at high sampling rates or fail to converge for low SNR and low dynamic range. We propose an algorithm to address these issue, that is based on the fact that the residual sequence, difference between the true samples and the modulo samples, is time-limited and has spectra beyond that of the true signal. We show that our approach results in the lowest error for a given sampling rate, dynamic range, and SNR compared among existing approaches.

\newpage

\bibliographystyle{IEEEtran}
\bibliography{US_biblios,refs}
\end{document}